\documentclass[sigconf,screen,authorversion,nonacm]{acmart}

\AtBeginDocument{%
  }

\copyrightyear{2025}
\acmYear{2025}
\setcopyright{cc}
\setcctype{by-nc-nd}
\acmDOI{XXXXXXX.XXXXXXX}

\usepackage{listings}
\usepackage{xcolor}
\usepackage{multirow}

\definecolor{codegray}{rgb}{0.5,0.5,0.5}
\definecolor{codepurple}{rgb}{0.58,0,0.82}
\definecolor{backcolour}{rgb}{0.98,0.98,0.98}

\lstdefinestyle{pythonStyle}{
  language=Python,
  basicstyle=\ttfamily\small,
  backgroundcolor=\color{backcolour},
  keywordstyle=\bfseries,
  stringstyle=\color{codepurple},
  commentstyle=\itshape\color{codegray},
  showstringspaces=false,
  columns=fullflexible,
  keepspaces=true,
  frame=single,
  rulecolor=\color{black},
  numbers=left,
  numberstyle=\tiny\color{codegray},
  numbersep=6pt,
  breaklines=true,
  breakatwhitespace=true,
  tabsize=4,
  captionpos=b
}

\lstnewenvironment{pythoncode}[1][]%
  {\lstset{style=pythonStyle,#1}}{}

\lstdefinestyle{cpp}{
  language=C++,
  basicstyle=\ttfamily\small,
  columns=fullflexible,
  breaklines=true,
  frame=single,
  numbers=left,
  numberstyle=\tiny,
  captionpos=b
}

\lstnewenvironment{cppcode}[1][]
  {\lstset{style=cpp,#1}}
  {}

\acmConference[Conference acronym '25]{Make sure to enter the correct
  conference title from your rights confirmation email}{Month DD--DD,
  2025}{City, Country}
\acmISBN{978-1-4503-XXXX-X/2025/08}

\title[Towards a Pythonic PIMPL]{The Opaque Pointer Design Pattern in Python: Towards a Pythonic PIMPL for Modularity, Encapsulation, and Stability}

\author{Antonios Saravanos}
\authornote{The first and second authors contributed equally to this paper.}
\orcid{0000-0002-6745-810X}
\affiliation{%
  \institution{New York University}
  \streetaddress{7 East 12th Street, Room 625B}
  \city{New York}
  \state{NY}
  \country{USA}
  \postcode{10003}
}
\email{saravanos@nyu.edu}

\author{John Pazarzis}
\authornotemark[1]
\affiliation{%
  \institution{Independent Researcher}
  \city{New York}
  \state{NY}
  \country{USA}
}

\author{Stavros Zervoudakis}
\affiliation{%
  \institution{New York University}
  \city{New York}
  \state{NY}
  \country{USA}
}

\author{Dongnanzi Zheng}
\affiliation{%
   \institution{New York University}
  \city{New York}
  \state{NY}
  \country{USA}
}

\renewcommand{\shortauthors}{Saravanos, et al.}



\begin{document}
\begin{abstract}
Python libraries often need to maintain a stable public API even as internal implementations evolve, gain new backends, or depend on heavy optional libraries. In Python, where internal objects are easy to inspect and import, users can come to rely on ``reachable internals'' that were never intended to be public, making refactoring risky and slowing long-term maintenance. This paper revisits the pointer-to-implementation (PIMPL) idiom from C++ and reinterprets it as a Pythonic pattern of opaque delegation: a small public object (or module) that delegates its behavior to a separate implementation object treated as internal. We situate this pattern within a broader taxonomy of encapsulation techniques in Python, relate it to existing practices such as module-level indirection, facade objects, and backend dispatch, and identify PIMPL-like structures already used in the standard library and the scientific Python ecosystem. We then show how a Pythonic PIMPL can be used in existing codebases to isolate heavy dependencies, support lazy imports, and enable runtime selection of alternative backends without changing the public API. Finally, we discuss the benefits and trade-offs of the approach and offer practical guidance on when the pattern is appropriate and how to apply it in large, long-lived Python libraries.
\end{abstract}

\maketitle

\section{Introduction}
Modern large-scale Python libraries increasingly need to balance rich functionality with long-term maintainability. As codebases grow, it becomes important to hide complex internal logic behind clean public interfaces, allowing internal components to evolve without disrupting users. At the same time, Python’s culture of openness and introspection, combined with its lack of enforced access control, makes it easy for implementation details to leak into what users perceive as the public API~\cite{Slatkin2015}. In the C++ world, the opaque pointer design pattern, known more commonly as the pointer to implementation (PIMPL) idiom~\cite{Coplien1992,SutterPimpl,CppreferencePImpl}, is a well-established technique for separating a class’s stable public interface from its changeable internals.

We should first clarify what we mean by a \emph{design pattern} and an \emph{idiom}. Following the usual definition in the literature~\cite{Gamma1994,bosch1998design}, a design pattern describes a recurring solution to a software-design problem at a level above any particular language. Examples include the facade, strategy, and bridge patterns. An \emph{idiom}, by contrast, is language specific. It captures a conventional way of realizing a design idea using a particular language’s syntax, semantics, and tooling. In this sense, the opaque pointer is a design pattern available in C and other languages with explicit pointers, whereas the C++ PIMPL is widely described as an idiom~\cite{Coplien1992}, as it specializes the opaque-pointer idea to C++ and the properties of that specific language. The PIMPL idiom also appears under other labels; some examples are listed in Table~\ref{tab:pimpl-names}.

\begin{table*}[t]
  \centering
  \renewcommand{\arraystretch}{1.20}
  \begin{tabular}{p{0.20\textwidth} p{0.43\textwidth} p{0.31\textwidth}}
    \toprule
    \textbf{Idiom} & \textbf{Brief definition} & \textbf{Origin/naming} \\
    \midrule
    Pointer to Implementation \\
    (PIMPL) &
      Pointer from an interface to a hidden implementation object. &
      Coined by Jeff Sumner, idiom described by Coplien~\cite{Coplien1992} and later Sutter~\cite{SutterPimpl,SutterJoyOfPimpls,SutterGotW100}. \\

    Opaque Pointer &
      C-style technique hiding representation behind an incomplete type and function API. &
      Used as a systems term, particularly for C abstract data types~\cite{Hanson1997,OpaquePointerWiki}. \\

    Handle--Body &
      Stable handle object separated from mutable body. &
      Attributed to Coplien~\cite{Coplien1992}. \\

    Envelope--Letter &
      Thin envelope that carries an internal ``letter'' where data and logic reside. &
      Introduced and named by Coplien~\cite{Coplien1992}. \\

    Compiler Firewall &
      Reduces dependencies by hiding private members behind an opaque type. &
      Popularized by Sutter~\cite{SutterExceptionalCpp,SutterJoyOfPimpls,SutterGotW100}. \\

    Cheshire Cat &
      Visible class can outlive or change hidden implementation. &
      Coined by Carolan, discussed by Meyers~\cite{Carolan1989,MeyersEffectiveCpp}. \\

    D-pointer &
      Qt-style technique using a dedicated \texttt{d\_ptr} member. &
      Credited to Gulbrandsen, documented in Qt/KDE binary-compat notes~\cite{KdeBinaryCompat,QtDPointer}. \\
    \bottomrule
  \end{tabular}
  \caption{Idioms for separating interface from implementation in C++.}
  \label{tab:pimpl-names}
\end{table*}

Although Python does not face the same compilation or binary-compatibility constraints that originally motivated PIMPL in C++~\cite{KdeBinaryCompat,Hanwell2023PIMPL}, many of the idiom’s architectural benefits apply directly to large Python systems. As Python has become foundational in scientific computing, machine learning, and high-performance infrastructure, developers increasingly face challenges such as managing heavy or optional dependencies, refactoring internal components without breaking downstream code, coordinating multiple backend implementations, and preserving stable public APIs across versions~\cite{Lowekamp2013SimpleITK,nep52,pandasInternalAPI,PEP399}.

This paper examines how the structural essence of the PIMPL idiom can be adapted to Python, aligning with Pythonic conventions and capabilities~\cite{OttingerPythonPimpl}. Rather than proposing a rigid design pattern, we reinterpret PIMPL conceptually as a form of opaque delegation. A public-facing object delegates its behavior to an internal implementation object, typically stored in a private attribute (e.g., \texttt{self.\_impl}) or hidden behind a module-level indirection layer. This approach preserves the separation of interface and implementation found in classical PIMPL.

Structuring components in this way offers several practical benefits. Heavy or optional dependencies can be isolated in the implementation object and imported lazily~\cite{lazy-loader,ScientificPythonSPEC1,PEP690,HRTLazyImports}. Multiple backend implementations can be selected dynamically based on configuration or runtime conditions and possibly complemented by the injection of control (IoC) pattern~\cite{PyPAGuidePlugins,EntryPointsSpec}. Internal algorithms and data structures can evolve behind a stable interface, reducing the risk of breaking changes. These capabilities are especially relevant for widely used, long-lived libraries that must evolve without sacrificing stability, performance, or extensibility.

This paper makes three contributions. First, it articulates a Pythonic PIMPL pattern that maps the classical PIMPL idiom onto Python’s object and module systems, including both class-based and module-level variants. Second, it relates this pattern to existing ecosystem practices (e.g., facade objects, backend selection, module-level indirection) and shows how they can be viewed as instances of a common opaque delegation structure. Third, it presents practical guidance and examples that demonstrate how and when to introduce a Pythonic PIMPL structure in real-world codebases, along with the benefits and trade-offs.

The remainder of this paper is organized as follows. Section~\ref{sec:background} reviews opaque pointers, opaque handles, and the classical C++ PIMPL idiom and contrasts them with Python’s approach to encapsulation, identifying similar structures already in use across the Python ecosystem. Section~\ref{sec:why-python-pimpl-full} then explains why Python might benefit from a PIMPL-style idiom, articulating the design pressures and concrete problems that motivate our proposal. Section~\ref{sec:pattern-overview} formalizes this recurring structure as the \emph{Pythonic PIMPL}, detailing its intent, applicability, structure, and architectural consequences. Section~\ref{sec:real-world} surveys real-world examples of PIMPL-like structures in the Python ecosystem and relates Pythonic PIMPL to existing design patterns and library idioms. Section~\ref{sec:discussion} discusses practical considerations and trade-offs in applying Pythonic PIMPL in large, long-lived Python codebases. Section~\ref{sec:conclusion} concludes the paper and outlines the limitations of our work, along with directions for future research.

\section{Background}
\label{sec:background}

This section situates our proposed Pythonic PIMPL pattern within existing language and ecosystem practices. We first review how opaque pointers, opaque handles, and the classical PIMPL idiom emerged in C and C++ to separate stable interfaces from volatile implementations~\cite{Hanson1997,OpaquePointerWiki,Coplien1992,Carolan1989,SutterPimpl}. We then contrast this with Python's more informal approach to encapsulation and consider PIMPL-like structures that already appear in the Python ecosystem.

At first glance, it may seem unnecessary, or even misleading, to speak of a
\emph{Pythonic PIMPL}. Python does not have explicit pointers, does
not rely on header files, and does not face the compilation and application binary interface (ABI)
constraints that originally motivated PIMPL in C++~\cite{KdeBinaryCompat,QtDPointer,SutterGotW100,MicrosoftLearn2025Pimpl,Hanwell2023PIMPL}. However, the primary architectural benefit of PIMPL is that it provides a deliberately narrow, stable interface boundary to shield clients from a more complex, volatile, or dependency-heavy implementation.

Python already has mechanisms that play the same architectural role,
even if they do not involve raw pointers. For example, many libraries
rely on module-level indirection, in which a public module reexports or
lazily resolves names while the real implementation lives in private
submodules (for instance, \texttt{socket}/\texttt{\_socket},
\texttt{pickle}/\texttt{\_pickle}, or SciPy's lazy-loading machinery)~\cite{PEP399,PythonDocsElementTree,PythonDocsSocket,PythonDocsPickle,ScientificPythonSPEC1,SciPyAPI,lazy-loader,PEP562,PEP690}.
Small public classes often act as  wrappers that hold an internal
implementation object (often in \texttt{self.\_impl}) and forward behavior
to it~\cite{OttingerPythonPimpl,Slatkin2015}. In other cases, frameworks organize their functionality through
backend or plugin dispatch layers, where a stable public API routes
operations to interchangeable backends (e.g., pure Python versus
accelerated,  local versus remote) selected at runtime or by configuration~\cite{PyPAGuidePlugins,EntryPointsSpec,nep52,pandasInternalAPI,SciPyFFTBackend,NumPyDispatch}.

Taken together, these practices form a recurring structural pattern:
Python developers use indirection to keep internal structure malleable
while presenting a clean, stable, well-curated public interface. What
Python lacks in pointer syntax, it compensates for with flexible module
boundaries, dynamic dispatch, descriptors, attribute-resolution hooks,
and language boundaries to native extensions. In this sense, a
Pythonic PIMPL is not a transliteration of C++ PIMPL, but a
language-appropriate idiom capturing a design that already exists
in the ecosystem~\cite{Gamma1994,bosch1998design,MeyersEffectiveCpp,Slatkin2015,PEP20}.

\subsection{Definitions and Relationships of Opaque Pointers, Handles, and PIMPL}
\label{sec:opaque-definitions}

Software systems have long relied on indirection to separate interface
from implementation. Two related low-level mechanisms, opaque pointers
and opaque handles, and the higher-level C++ PIMPL idiom embody this
strategy at different levels. They share the motivation of hiding
the implementation and reducing coupling, but they arise in different historical
contexts and are shaped by language features.

\subsubsection{Opaque Pointers}

An opaque pointer is a pointer to an incomplete type whose internal
representation is intentionally hidden from clients~\cite{Hanson1997,OpaquePointerWiki}. In C, this pattern
is achieved by forward-declaring a \texttt{struct} in a header (for
example, \texttt{typedef struct my\_struct my\_struct;}) while defining the
struct's fields only in a private source file. Client code manipulates
such objects exclusively through pointers and API functions that
accept or return them, never through direct access to their fields. Opaque pointers
enable strict encapsulation, representation independence, reduced
compile-time dependencies, and stable ABIs because client code does not
depend on the structure's layout~\cite{Hanson1997}.

\subsubsection{Opaque Handles}

Opaque handles generalize the opaque-pointer concept by abstracting the
reference mechanism itself~\cite{OpaquePointerWiki}. A handle may be a pointer, an integer, an
index into an internal table, or some other token. Operating systems and
GPU APIs make extensive use of such handles to reference resources that
must be managed behind a protection boundary: process identifiers, file
descriptors, synchronization primitives, GPU buffers, pipelines, and
devices. In these contexts, the term \emph{handle} expresses the ideas of resource
ownership, capability control, and lifetime management. Internally, the
system maintains tables that map handles to concrete resources, and
clients invoke operations that consume or produce handles rather than
inspecting representations directly. Opaque pointers and opaque handles thus form a family of
indirection-based encapsulation mechanisms in C and systems-level APIs, supporting resource
isolation and hidden representation. The C++ PIMPL idiom, discussed in
Section~\ref{sec:background-cpp-pimpl}, specializes the same idea for
classes and library interfaces~\cite{Coplien1992,SutterPimpl,CppreferencePImpl}.

\subsection{The PIMPL Idiom in C++}
\label{sec:background-cpp-pimpl}

The PIMPL idiom is a C++ technique for
separating a class’s stable public interface (in a header) from its changeable
private representation (in a source file). It is used to limit dependency
propagation (a “compiler firewall”) and, in compiled library settings, to help
preserve ABI stability by keeping the header-visible class layout fixed~\cite{Coplien1992,SutterPimpl,SutterGotW100,CppreferencePImpl}. The header forward-declares a private \texttt{Impl} type and stores only an
opaque pointer to it (commonly \texttt{std::unique\_ptr<Impl>}); the full
\texttt{Impl} definition lives in the \texttt{.cpp} file. With \texttt{unique\_ptr},
the public destructor is typically defined \emph{out-of-line} so that \texttt{Impl}
is complete at the deletion point~\cite{CppreferencePImpl}. Listing~\ref{lst:cpp-pimpl-min}
shows the canonical structure: a header-visible \texttt{Widget} that owns only
\texttt{impl\_}, while all state and behavior live in \texttt{Widget::Impl} in
\texttt{widget.cpp}.

\begin{cppcode}[caption={Minimal PIMPL structure in C++ (header + source)}, label={lst:cpp-pimpl-min}]
/* widget.h */

#ifndef WIDGET_H
#define WIDGET_H

enum State {
    STATE_IDLE = 0,
    STATE_RUNNING = 1
};

class WidgetImpl;

class Widget final {
public:
  Widget();
  ~Widget();

  void start();
  void stop();
  State  get_state() const;

private:
  Widget(const Widget&);
  Widget& operator=(const Widget&);
  Widget(Widget&&);

private:
  WidgetImpl* m_pImpl;
};

#endif  // WIDGET_H

/* widget.cpp */

#include "widget.h"
#include <stdexcept>

class WidgetImpl {
    public:
      WidgetImpl() : _state(STATE_IDLE) {
      }

      void start() {
          if (_state == STATE_IDLE) {
              _state = STATE_RUNNING;
          }
          else {
              // Already running
              throw std::runtime_error("Widget is already running");
          }
      }

      void stop() {
          if (_state == STATE_RUNNING) {
              _state = STATE_IDLE;
          }
          else {
              throw std::runtime_error("Widget is already stopped");
          }
      }

    State get_state() const {
        return _state;
    }

    private:
      State _state;
};

Widget::Widget() : m_pImpl(new WidgetImpl) {

}

Widget::~Widget() {
    if (m_pImpl != nullptr) {
        delete m_pImpl;
        m_pImpl = nullptr;
    }
}

void Widget::start() {
    m_pImpl->start();
}

void Widget::stop() {
    m_pImpl->stop();
}

State Widget::get_state() const {
    return m_pImpl->get_state();
}
\end{cppcode}

Because client code sees only the pointer member, implementation headers and
the private data layout are hidden; changing \texttt{Impl} often avoids downstream
rebuilds and can help keep an exported class’s size and layout stable across library
versions~\cite{KdeBinaryCompat,QtDPointer,Hanwell2023PIMPL}. The trade-offs are
extra indirection and usually one heap allocation per object (plus potential
loss of inlining and less direct debugging)~\cite{SutterPimpl,Hanwell2023PIMPL}.
PIMPL is not a universal ABI shield: changes to exported signatures, inline
definitions, or virtual interfaces can still break ABI even if private data are
hidden~\cite{CppreferencePImpl}.

\subsection{Encapsulation in Python}
\label{sec:background-python-encapsulation}

Python approaches encapsulation differently from statically typed or compiled languages such as C++ or Java. It favors readability, simplicity, and developer freedom over rigid enforcement of access control. The interpreter does not support private fields or access modifiers in the traditional sense; any attribute can, in principle, be accessed or mutated from outside the class or module. Instead, intent is communicated through naming conventions and social norms, often summarized by the phrase “we are all consenting adults here”.

A single leading underscore (e.g., \texttt{\_helper}) signals that a function, variable, or attribute is intended for internal use. This affects wildcard imports (i.e., \texttt{from module import *}) by excluding such names, but it imposes no actual restrictions on access. As a stronger form of encapsulation, Python supports name mangling via double underscores. When a class attribute or method is prefixed with two leading underscores (e.g., \texttt{\_\_internal}), the interpreter rewrites the attribute name to include the class name internally (e.g., \texttt{\_MyClass\_\_internal}). This makes it harder to access the attribute directly from outside the class and reduces the chance of accidental name clashes in subclasses.

In effect, name mangling provides a lightweight mechanism for discouraging external access, but it does not offer true privacy or enforcement. Developers can still access mangled names deliberately if they know the internal naming convention, and tools like \texttt{dir()} or reflection APIs make such attributes visible. Nonetheless, this mechanism can reduce the likelihood of accidental access and reinforces the notion that certain components are internal implementation details.

Modules may also use the \texttt{\_\_all\_\_} variable to define a curated public API, even though all contents of the module remain technically importable. This flexibility underpins common Python idioms, such as monkey patching, runtime substitution of components, deep introspection, and ad hoc dependency injection. It also simplifies testing, since internal objects remain accessible. As a result, Python rarely requires or uses formal dependency injection frameworks as seen in languages with stricter encapsulation.

However, as Python projects scale, the absence of hard boundaries can create fragility and maintenance pressure. Users may come to rely on undocumented or “semi-private” attributes (i.e., names, attributes, or constructs that are intended to be private but are not actually enforced by the language runtime) that were never intended to be part of the public interface. This makes refactoring internals risky, as it can inadvertently break downstream code. Major scientific libraries now explicitly warn that internal submodules and names may change without notice and should not be relied upon~\cite{pandasInternalAPI}. NumPy’s work toward its 2.0 release, for example, includes efforts to clarify public versus private APIs and reduce ambiguous “reachable internals” \cite{nep52}.

This tension between openness and stability is central to the architecture of large Python codebases. On one hand, Python’s dynamic nature and introspection-friendly culture provide powerful tools for rapid development and experimentation. On the other hand, long-lived libraries increasingly require structural boundaries to separate stable public APIs from internal implementations. In this architectural space, patterns that approximate encapsulation through indirection, such as Pythonic PIMPL, offer a disciplined yet flexible compromise.

Although Python lacks strict enforcement of access control, a range of idioms and mechanisms have evolved to provide practical encapsulation. These strategies differ in their reliance on naming conventions, module structure, runtime delegation, or execution boundaries (i.e., runtime or interpreter-level barriers that naturally hide data, even if Python’s syntax does not). We organize them into four broad categories.

The first category, which we call stylistic encapsulation, relies on social conventions rather than any enforced mechanism. A leading single underscore in an attribute or module name (e.g., \texttt{\_helper}) signals to users that the name is internal. This affects wildcard imports but imposes no restriction on access. Double underscores (e.g., \texttt{\_\_name}) trigger name mangling, which makes accidental access less likely by rewriting names internally (e.g., \texttt{\_\_x} becomes \texttt{\_ClassName\_\_x}), but this too is bypassable. Some developers may at times mark modules or subpackages as private by naming them with a leading underscore (e.g., \texttt{package.\_internal}), signaling non-public status without preventing imports. These conventions are lightweight and developer-friendly but provide no actual hiding: the implementation remains accessible and mutable.

The second category centers on namespace and module boundary encapsulation. In this strategy, modules define a curated public API using the \texttt{\_\_all\_\_} variable to specify which symbols are intended for export. Implementation code may be placed in private submodules, such as \texttt{\_core} or \texttt{\_impl}, from which a top-level package or module reexports selected names to define a stable interface. This allows large packages to expose a clean and hierarchical API surface. Although this approach provides structural clarity and improves documentation, it does not prevent direct access to internal modules or names, as everything remains importable given the correct path.

The third category is encapsulation by indirection, which includes design patterns that explicitly separate interface from implementation using delegation. This includes public facade classes that forward behavior to internal helper classes, typically stored in attributes such as \texttt{self.\_impl}. The indirection may also take the form of proxies or wrappers that encapsulate behavior and restrict direct access to internal state. This is the category in which the Pythonic PIMPL pattern naturally resides. By maintaining an internal implementation object that is not exposed as part of the public API, Python developers can achieve a meaningful separation of concerns and reduce the surface area exposed to users. Although this strategy is not enforced by the language runtime and can be bypassed via introspection, it is typically sufficient in practice and helps maintain modular and testable code.

The fourth and final category involves encapsulation by execution environment or language boundary. This is most commonly seen in mixed-language systems that hold internal state in C extensions or other native code not directly accessible from Python. Modules like \texttt{socket} and \texttt{pickle} exemplify this strategy: they expose a public Python interface that forwards to a private C implementation (e.g., \texttt{\_socket}, \texttt{\_pickle}). Mechanisms such as \texttt{PyCapsule} allow opaque C pointers to be passed through Python without exposing their structure.

\subsection{Implicit PIMPL-Like Patterns in Python}
\label{sec:pimpl-like-idioms}

Although the term ``PIMPL'' is not commonly used in the Python community, a
number of established practices closely resemble its intent of hiding volatile or
heavy implementation details behind a small, stable interface~\cite{PEP399,PythonDocsElementTree,PyPAGuidePlugins}.

Early community discussions explicitly asked whether PIMPL makes sense in
Python. For example, Ottinger~\cite{OttingerPythonPimpl} experimented with a facade object that delegates all work to a hidden implementation, primarily to enable easy swapping of implementations in tests. Follow-up threads on community forums and news aggregators have argued that Python's dynamic nature already provides most of PIMPL's benefits without additional ceremony, and some contributors concluded that a dedicated PIMPL pattern has limited real-world value beyond ordinary composition and delegation~\cite{HNPimpl2024}. These explorations reveal both recurring interest in an opaque implementation surface and a degree of skepticism about whether it deserves to be named as a separate pattern. Part of this skepticism reflects broader differences in how language communities approach design patterns. Developers coming from dynamic, mixed-paradigm languages often rely on ad hoc composition and runtime flexibility, whereas communities rooted in more strictly object-oriented ecosystems tend to think in terms of named patterns and explicit interface--implementation separations.

At the same time, several widely used libraries employ structures that are like PIMPL in everything but name. Widely used patterns echo the same intent. Delegating PIMPL objects route calls to backend implementations, public modules act as proxies or frontends to hidden C extensions~\cite{PEP399}, and module-level indirection enables lazy loading and namespace control~\cite{lazy-loader,ScientificPythonSPEC1,PEP690,HRTLazyImports}. In the scientific Python ecosystem, such indirection has been adopted to reduce import time and decouple public APIs from internal package structure~\cite{nep52,pandasInternalAPI}, even when it is not described explicitly in PIMPL terms.

\section{Why Python Might Need a PIMPL Pattern}
\label{sec:why-python-pimpl-full}

Large Python projects often contain components with complex logic,
optional accelerators, or platform-specific behavior~\cite{PEP399,PythonDocsElementTree}.
If public interfaces and internal machinery are intermixed,
implementation details tend to leak into the visible API, making
evolution risky and refactoring harder.
Consider a data processing library that exposes a public class
\texttt{DataProcessor}.
Internally, different implementations may exist, such as a pure
Python version, a NumPy-based accelerator, or a platform-specific
backend.
A PIMPL-style arrangement lets the public \texttt{DataProcessor}
select an appropriate implementation, store it in \texttt{self.\_impl},
and delegate all user-facing methods to that object.
Users see only the stable interface, while maintainers can update or
replace the implementation without changing call sites, in the spirit
of the C++ PIMPL idiom~\cite{CppreferencePImpl,OttingerPythonPimpl}.

Even though Python does not enforce strict privacy and relies instead
on naming conventions for public vs.\ non-public
names~\cite{PEP8,RealPythonUnderscores},
this separation still provides concrete benefits. It keeps heavy
dependencies and complex logic in dedicated classes or modules,
enables lazy imports, and keeps the public API concise and
predictable~\cite{RealPythonAll,PEP562,ScientificPythonSPEC1}.

This separation offers several practical benefits for large or
long-lived Python libraries:

\begin{enumerate}
  \item \textbf{Stable public APIs.}
        A small public interface isolates users from internal churn,
        clarifies what is supported as part of the public surface,
        and matches community guidance to distinguish public and internal names explicitly in modules and packages~\cite{PEP8,RealPythonAll,RealPythonUnderscores}.

  \item \textbf{Substitute implementations and testing.}
        Backend swapping becomes an explicit design choice via a
        delegating layer, rather than ad hoc globals or monkey
        patching.
        This is similar in spirit to plugin architectures that load
        implementations through entry points or dynamic discovery
        mechanisms~\cite{PyPAGuidePlugins,EntryPointsSpec}.

  \item \textbf{Performance (startup and memory).}
        Lazy imports in the implementation layer reduce import time
        and memory footprint, for example, through module-level
        indirection, PEP~562-style attribute loading, and library
        patterns for lazy submodule
        exposure~\cite{PEP562,ScientificPythonSPEC1}.
        Interpreter-level lazy-import experiments have quantified potential improvements in startup times and memory usage, while demonstrating the importance of carefully managing side effects and semantics~\cite{PEP690,HRTLazyImports}.

  \item \textbf{Cross-language mindshare.}
        Naming the idiom \emph{Pythonic PIMPL} gives mixed C++ and
        Python teams a shared term for a stable interface with an
        opaque implementation, echoing the established PIMPL pattern
        in C++ and similar delegation tricks already used in
        Python~\cite{CppreferencePImpl,OttingerPythonPimpl}.

  \item \textbf{Reducing user error.}
        Clear boundaries between the public interface and internal
        implementation discourage accidental reliance on internals
        and make documentation easier to align with reality, in accordance
        with guidance to keep non-public names and implementation
        details out of the documented API
        surface~\cite{PEP8,RealPythonAll,RealPythonUnderscores}.
\end{enumerate}

Not every project needs this separation, and small libraries may
prefer simpler approaches. For widely used, long-lived components, however, a thin indirection layer can unlock safer evolution without sacrificing Python's flexibility, much as lazy-loading schemes are recommended primarily for large libraries where import overhead is a real concern~\cite{ScientificPythonSPEC1}.

\section{The Pythonic PIMPL Pattern}
\label{sec:pattern-overview}

The Pythonic PIMPL pattern adapts the core idea behind opaque pointers and the C++ PIMPL idiom, encapsulation through indirection, to Python’s object model. Although Python lacks pointers and does not enforce visibility constraints, it supports dynamic features that make it possible to create a disciplined separation between interface and implementation. A Pythonic PIMPL formalizes this separation using a lightweight delegation structure: a stable public object whose behavior is implemented by a hidden internal object. In this model, the public object is conceptually analogous to the “handle” or “frontend,” and the private object serves as the “implementation.”

This pattern is motivated by a desire to improve modularity, encapsulation, and API stability in large or evolving codebases. Isolating implementation details behind a private indirection layer helps reduce coupling and simplifies long-term maintenance. Encapsulation is achieved through structural conventions that discourage reliance on internal components. Most importantly, Pythonic PIMPL makes it easier to refactor, optimize, or even replace entire backends without breaking user-facing code. It achieves these goals while remaining faithful to Python’s emphasis on clarity, runtime flexibility, and maintainable design. Table~\ref{tab:pythonic-pimpl-summary} provides a concise summary of the pattern in a catalog style.

\begin{table}[h]
\centering
\caption{Pattern Summary: Pythonic PIMPL}
\label{tab:pythonic-pimpl-summary}
\begin{tabular}{p{0.27\linewidth} p{0.63\linewidth}}
\toprule
\textbf{Name} & \textbf{Pythonic PIMPL} \\

\midrule
\textbf{Also Known As} & Python Private Implementation Idiom; Envelope--Letter Idiom (Python); Handle--Body Idiom (Python) \\

\midrule
\textbf{Intent} & To decouple a stable public interface from a volatile or complex implementation, enabling internal evolution while preserving API stability in Python libraries. \\

\midrule
\textbf{Motivation} & Large Python frameworks require long-lived, stable APIs while allowing internal refactoring, alternative backends, lazy imports, and heavy-dependency isolation. Pythonic PIMPL encapsulates volatility behind a lightweight interface object. \\

\midrule
\textbf{Applicability} & Use when: (1) the implementation is likely to change; (2) optional or heavy dependencies should be hidden; (3) multiple backend implementations exist; (4) a stable API must be preserved for external users; (5) you want to avoid dependency on third-parties (i.e., vendor lock-in). \\

\midrule
\textbf{Structure} & A public \emph{Interface Object} contains a private reference (typically \texttt{\_impl}) to an \emph{Implementation Object}. All public methods delegate to the implementation. This may be class-based or module-level (via PEP~562 or hidden backend modules). \\

\midrule
\textbf{Participants} &
\textbf{Client}: consumes the public API. \newline
\textbf{Interface}: exposes stable public methods. \newline
\textbf{Implementation}: contains hidden logic, heavy imports, or backend-specific code. \\

\midrule
\textbf{Consequences (Benefits)} &
Improves modularity; isolates internal churn; supports multiple backends; reduces import cost via lazy loading; provides a clean API boundary; enhances testability. \\

\midrule
\textbf{Consequences (Liabilities)} &
Adds delegation overhead; increases boilerplate; may introduce debugging indirection; risks over-engineering if used for small components. \\

\midrule
\textbf{Known Uses} &
Python stdlib (e.g., \texttt{socket}/\texttt{\_socket}, \texttt{pickle}/\texttt{\_pickle}); scientific libraries; GUI toolkits; plugin-based frameworks. \\

\midrule
\textbf{Related Patterns} &
C++ PIMPL; Bridge; Proxy; Facade; Strategy; Abstract Factory (for implementation selection). \\

\bottomrule
\end{tabular}
\end{table}

\subsection{Name and Also Known As}

We call this pattern the \emph{Pythonic PIMPL}. The name reflects its origin in the classical PIMPL idiom reviewed in Section~\ref{sec:background-cpp-pimpl}, but emphasizes a formulation that fits Python’s object and module systems. In analogy with the C++ literature, it can also be viewed as a Python-specialized form of the handle–body or envelope–letter idioms.

\subsection{Intent}

The intent of the Pythonic PIMPL pattern is to separate a module or class interface from its internal implementation in order to reduce coupling, hide details that are likely to change, and minimize dependencies visible to users. Whereas C++ uses PIMPL primarily to address compilation and binary compatibility constraints, the same structural idea can be applied in Python to achieve clarity, modularity, and insulation from internal changes.

Concretely, a lightweight \emph{interface} object forwards operations to an internal \emph{implementation} object stored in an attribute such as \texttt{self.\_impl} or hidden behind a module-level indirection layer. This preserves the outward shape of the API while permitting the internal logic, data structures, and dependencies to evolve freely.

\subsection{Motivation}
\label{sec:why-python-pimpl}

The full motivation for why Python benefits from a PIMPL-style idiom is
presented in Section~\ref{sec:why-python-pimpl-full}.
This subsection serves only as a brief summary within the pattern catalog,
mirroring the “Motivation” entry found in classical pattern descriptions. In short, large Python libraries frequently need to preserve stable public APIs
while allowing their internal logic, data structures, and dependencies to evolve.
Implementations may involve optional accelerators, platform-specific backends,
or heavy imports that should not appear at the top-level public interface.
A Pythonic PIMPL structure protects the public API from these internal details
and enables safe evolution over time.




\subsection{Applicability}

A Pythonic PIMPL structure is useful in several recurring situations. It is a natural fit for large or complex components whose implementation details would otherwise clutter the public API, and for externally visible APIs whose internals are expected to change as algorithms, data structures, or dependencies evolve. It also helps when the implementation depends on optional or heavy libraries that should not be imported eagerly. Those imports and related logic are moved into an internal implementation object. The same structure works well when multiple interchangeable implementations exist, such as accelerated variants and pure-Python fallbacks, or platform-specific versions, since the public interface can select and construct the appropriate backend while preserving a single, stable interface. In larger organizations, the interface/implementation split can also support separation of concerns, with one team owning the public interface and another maintaining backend implementations behind it.

For small, self-contained classes used only within a single module, this extra layer is usually unnecessary and may constitute overengineering. The pattern is most valuable in long-lived libraries and frameworks, which are common in scientific computing, machine learning, data processing, and networked systems, where APIs must remain stable even as internal structures and dependencies change over time.

\subsection{Canonical Example}
\label{sec:canonical-example}

Listings 2 and 3 illustrate a complete Pythonic PIMPL arrangement using a public module and a private implementation module:

\begin{pythoncode}[caption={Public \texttt{Widget} API module delegating to an internal implementation},
                   label={lst:widget-api}]
"""widget.py -- Public API module"""

import impl._widget as _widget

_WidgetImpl = _widget._WidgetImpl

class Widget:
    def __init__(self, config):
        self._impl = _WidgetImpl(config)

    def start(self):
        return self._impl.start()

    def stop(self):
        return self._impl.stop()

    def status(self):
        return self._impl.status()
\end{pythoncode}

\begin{pythoncode}[caption={Private \texttt{\_WidgetImpl} class used internally by \texttt{Widget}},
                   label={lst:widget-impl}]
"""impl._widget.py -- Private implementation module"""

class _WidgetImpl:
    def __init__(self, config):
        self._state = "initialized"
        self._config = config

    def start(self):
        self._state = "running"

    def stop(self):
        self._state = "stopped"

    def status(self):
        return {
            "state": self._state,
            "config": dict(self._config),
        }
\end{pythoncode}

In this example, the \texttt{\_WidgetImpl} class remains entirely private. It is never re-exported by the public module and does not appear directly in the documented API. Users of \texttt{Widget} interact only with its stable interface methods \texttt{start}, \texttt{stop}, and \texttt{status}. Internally, these methods simply delegate to the implementation object stored in \texttt{self.\_impl}.

Because callers never depend on the layout or existence of the \texttt{\_WidgetImpl}, the implementation can be extensively rewritten, data structures changed, algorithms replaced, helper functions reorganized, without breaking user code, provided that the public \texttt{Widget} methods preserve their semantics. The structure is directly analogous to the C++ PIMPL idiom. Specifically, a stable handle object owning an opaque implementation object behind an indirection barrier.

In real Python libraries, this idea is often applied in three steps. First, the public module shows only a small set of supported names. Second, a public class acts like a thin wrapper and forwards calls to a hidden object (for example, self.\_impl). Third, heavy or optional dependencies are kept inside the hidden code and imported only when needed. This keeps the public API stable while the internal code can change safely.

\subsection{Participants and Responsibilities}
\label{sec:pimpl-participants}

\begin{figure*}
    \centering
    \includegraphics[width=0.7\textwidth]{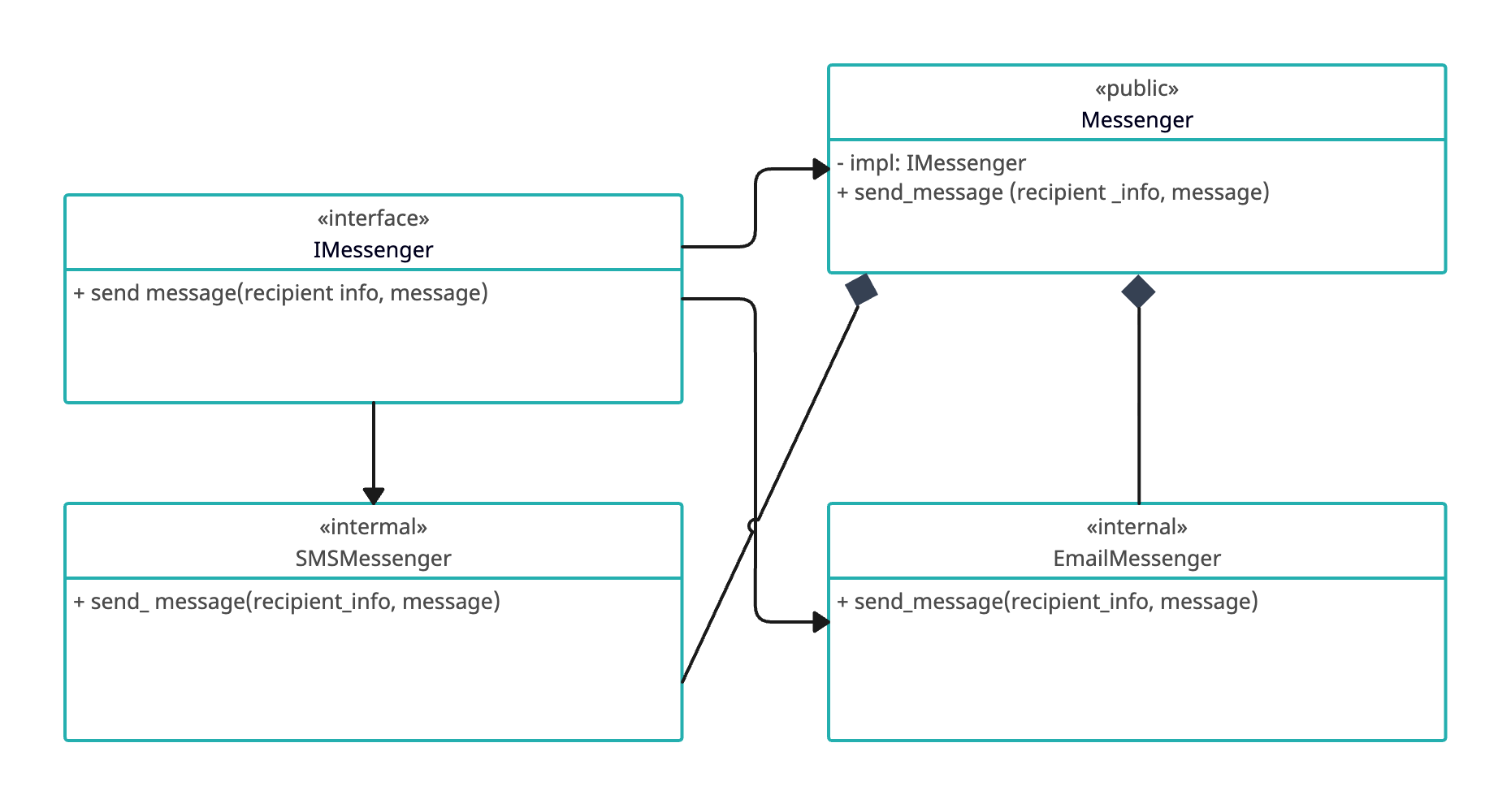}
    \caption{Interface–implementation separation in the Pythonic PIMPL pattern.}
    \label{fig:placeholder}
\end{figure*}

This subsection describes the conceptual roles that recur in a Pythonic PIMPL
arrangement and clarifies the responsibilities associated with each role.
Although Python does not enforce interface boundaries in the same way as C++,
the same architectural separation applies. Client code should depend on a
small, stable public surface, while the underlying implementation remains free
to evolve. Figure~\ref{fig:placeholder} illustrates this separation, and
Listings~\ref{lst:imessenger-interface} and~\ref{lst:messenger-class} provide a
concrete example in which a public PIMPL delegates to a runtime-selected
backend.

\subsubsection{Interface}
The interface defines the contract that all concrete implementations must
satisfy. It should remain small and stable: it declares method signatures and
documents semantics, but does not contain operational logic. To preserve
encapsulation, the interface should avoid dependencies on concrete implementation
classes, and its public signatures should use either built-in types or types that
are themselves part of the documented public API. In our example,
\texttt{IMessenger} plays this role, as illustrated in Listing~\ref{lst:imessenger-interface}.

\subsubsection{Classes derived from the interface}
Concrete classes implement the interface contract and provide the actual
behavior. These implementations may depend on lower-level helpers and
backend-specific code, and they should encapsulate the operational logic that is not intended to be exposed at the public layer. Heavy or optional dependencies are best
isolated here (or in private implementation modules imported by these classes),
so that importing the public API does not implicitly pull in expensive
dependencies. In Listing~\ref{lst:messenger-class}, the email and SMS backends
are examples of such concrete implementers that remain opaque to clients.

\subsubsection{Object factory}
The factory \cite{gfggofdesignpatterns2025} is responsible for selecting and creating the appropriate concrete implementation before returning an instance that is treated as the
interface type. It typically accepts configuration parameters that influence the
selection (e.g., platform, resource availability, user preferences). In Python
this factory role may appear as an explicit factory function (e.g.,
\texttt{create\_messenger(...)}). The \texttt{bind(...)} method in Listing~\ref{lst:messenger-class}
illustrates this selection step by choosing a backend at runtime while keeping
the public API stable.

\subsubsection{Client code}
Client code obtains an instance through the factory and interacts with it exclusively via the interface methods. This
keeps callers insulated from implementation details and allows implementations to
be swapped or refactored without requiring changes at call sites. In the example
of Listing~\ref{lst:messenger-class}, clients call \texttt{send\_message(...)}
on the interface that is exposed by PIMPL, while the concrete backend remains hidden, matching the separation
depicted in Figure~\ref{fig:placeholder}.\\

As an example, consider a \texttt{Messenger} (sender) class that implements a generic
interface like \texttt{IMessenger} and allows the user to send messages by email or SMS without exposing the implementation details.

\begin{pythoncode}[caption={Interface definition for a messenger service},
                   label={lst:imessenger-interface}]
"""The interface definition for a messenger service."""

import abc

class IMessenger(abc.ABC):
    """The interface for a messenger service."""

    @abc.abstractmethod
    def send_message(self, recipient, message):
        """Send a message to a recipient."""
\end{pythoncode}

\begin{pythoncode}[caption={Delegating \texttt{Messenger} class with runtime backend binding},
                   label={lst:messenger-class}]
"""A simple messenger service implementation."""

import messenger_interface
import impl.email_messenger as email_messenger
import impl.SMSMessenger as sms_messenger

class Messenger(messenger_interface.IMessenger):
    """Implements IMessenger for Email and SMS"""
    
    _pimpl = None

    def bind(self, messenger_type):
        """Bind the private implementation instance."""
        self._pimpl = None
        if messenger_type.lower() == "email":
            self._pimpl = email_messenger.EmailMessenger()
        elif messenger_type.lower() == "sms":
            self._pimpl = sms_messenger.SMSMessenger()
        else:
            raise RuntimeError

    def send_message(self, recipient, message):
        """Send a message to a recipient."""
        if not self._pimpl:
            raise RuntimeError
        if self._pimpl:
            self._pimpl.send_message(recipient, message)
\end{pythoncode}

In summary, Pythonic PIMPL is an interface–implementation split using explicit indirection. Clients call the stable interface, the wrapper selects a backend, and the wrapper forwards calls to a hidden implementation that can evolve without changes to the public API.

\subsection{Benefits and Trade-Offs}

The main benefits of the Pythonic PIMPL idiom are that it improves modularity, helps preserve API stability, supports backend flexibility, gives finer control over performance, and enhances testability~\cite{SutterPimpl,Hanwell2023PIMPL,PEP399,nep52,pandasInternalAPI}. By grouping implementation code into focused implementation classes or modules, the idiom keeps public interfaces smaller and easier to understand. Because internal changes occur behind the same hidden implementation layer~ \cite{pycaCryptographyBaseMain}, maintainers can often refactor or extend the implementation without altering the public interface~\cite{MeyersEffectiveCpp,SutterPimpl}. Multiple implementations can coexist behind the same frontend while presenting a single stable API to users. Isolating heavy or optional dependencies inside the implementation layer also allows them to be imported lazily and kept off the public module import path~\cite{lazy-loader,ScientificPythonSPEC1,PEP690,HRTLazyImports}. Finally, tests can inject mock or simplified implementations while exercising the same public interface, making it easier to test edge cases and failure modes~\cite{OttingerPythonPimpl}. Keeping in mind that object-oriented programming purists may object to exposing or mocking private details to the user.

The pattern also has some drawbacks. Each call passes through a delegating layer, introducing a small amount of runtime overhead and additional indirection that can complicate debugging~\cite{SutterPimpl,Hanwell2023PIMPL}. Naive applications of the pattern can add boilerplate, since many public methods may simply forward to similarly named methods on the implementation object unless helper utilities or metaprogramming are used~\cite{FilipekPimpl,SutterExceptionalCpp}. For simple components, this additional layer may amount to over-engineering, adding structure without delivering clear benefits. Moreover, because Python cannot truly prevent users from accessing attributes such as \texttt{\_impl}, the encapsulation boundary is partly conventional rather than enforced~\cite{PEP8,RealPythonUnderscores,PEP562}.

\section{Known Uses and Related Patterns}
\label{sec:real-world}

Several widely used libraries already employ PIMPL-like structures. The Python standard library often exposes a high-level Python module backed by a private C extension (e.g., \texttt{socket}/\texttt{\_socket}, \texttt{pickle}/\texttt{\_pickle}). Scientific libraries and plugin-based frameworks use PIMPL-like objects and module proxies to hide backend selection, heavy dependencies, and evolving internal package layouts.

Structurally, the Pythonic PIMPL idiom is closely related to the classical C++ PIMPL idiom, as well as the bridge, proxy, and facade patterns. Compared with these, its defining feature is the explicit use of an opaque implementation handle (often \texttt{\_impl}) in a Python context, with an emphasis on API stability, lazy loading, and backend swapping in large libraries.

Although the term Pythonic PIMPL is not widely used, several prominent
libraries have independently converged on structures that closely match the
pattern described in Section~\ref{sec:pattern-overview}. In each case, a small,
stable public interface delegates to an opaque implementation that can evolve
more freely. This section briefly surveys three representative families of
examples.

\subsection{Pythonic PIMPL Implementations Over Private C Extensions}

A long-standing idiom in the Python standard library is to expose a Python
module as the public API while placing performance-critical logic in a private
C extension module. For example, the \texttt{socket} module is documented as a
``wrapper module for \texttt{\_socket}, providing some additional facilities
implemented in Python''. The public module reexports selected classes,
functions, and constants from \texttt{\_socket} and augments them with
higher-level helpers, while the C extension remains an internal detail of the
CPython implementation. Similar arrangements exist for \texttt{pickle} and
\texttt{\_pickle} (formerly \texttt{cPickle}), \texttt{json} and
\texttt{\_json}, and several other libraries.

Structurally, these modules implement a Pythonic PIMPL at the module
level. The public module plays the role of the interface object, with a
stable, documented surface. The private C extension is the implementation
object, hidden behind the module boundary and free to change as long as it
preserves the exported contract. Users write against \texttt{socket.socket} or
\texttt{pickle.dump} and remain insulated from changes to the internal C data
structures or helper functions.

\subsection{Lazy Public Namespaces in Scientific Python}

Scientific Python projects have adopted a related idiom to manage import-time
cost and evolving package structure. Scientific Python SPEC~1 and associated
tooling such as \texttt{lazy\_loader} recommend presenting a flat, curated
public namespace while lazily importing heavy submodules on first use, often
via PEP~562's module-level \texttt{\_\_getattr\_\_} hook~\cite{PEP562}. Libraries such as SciPy, scikit-image, and NetworkX, have adopted variants of this pattern. A simplified example illustrates the structure:

\begin{pythoncode}[caption={Lazy public namespace using module-level \texttt{\_\_getattr\_\_}},
                   label={lst:lazy-namespace}]
"""public module: mylib/__init__.py"""

import importlib

__all__ = ["array", "read_image"]

def __getattr__(name):
    if name == "array":
        from ._core import array
        return array
    if name == "read_image":
        from ._io import read_image
        return read_image
    raise AttributeError(name)
\end{pythoncode}

From the user's perspective, \texttt{mylib.array} and \texttt{mylib.read\_image}
are stable, top-level entry points. Internally, those names are resolved at
runtime to functions defined in private submodules such as \texttt{mylib.\_core}
and \texttt{mylib.\_io}. The public module acts as an interface object. It
defines the set of supported names and their semantics, while the implementation
modules can be reorganized, split, or lazily imported without breaking callers.

This design matches the Pythonic PIMPL template, with the module itself holding
an implicit pointer to its implementation: the logic for each public name lives
elsewhere and is only revealed through the indirection layer.

\subsection{Hidden Cores and Accelerators in Large Libraries}

Some large third-party libraries also distinguish between a stable public surface
and more volatile internal cores. NumPy's work toward its 2.0 release, for
example, includes renaming the internal \texttt{numpy.core} package to
\texttt{numpy.\_core} and clarifying that it is not part of the public API
contract~\cite{nep52,quansight-numpy2}. The public \texttt{numpy}
namespace exposes a relatively small set of documented functions and
subpackages, while \texttt{numpy.\_core} provides the implementation of arrays
and ufuncs (which is a term specific to NumPy; see \cite{numpy-ufuncs-docs}) and is allowed to change more aggressively. Pandas follows a similar pattern. High-level classes such as \texttt{DataFrame} and \texttt{Series} form the documented interface, whereas performance-critical operations are implemented in internal Cython/C extension modules under \texttt{pandas.\_libs}. These extensions are tightly coupled to
the library's internals but are not intended as public extension points~\cite{pandasInternalAPI}. In both cases, the public Python objects form a stable interface that owns an
opaque, non-API implementation. The precise location, shape, and language of
the implementation can change (e.g., reorganizing \texttt{\_core}, adding or
removing C accelerators) without altering the public signatures or import paths
that users rely on~\cite{nep52,pandasInternalAPI}.

\subsection{The Pythonic PIMPL Idiom in Practice}

These examples suggest that a Pythonic analogue of PIMPL has emerged organically
across Python's ecosystem. Standard library modules wrap private C extensions;
scientific libraries use module-level indirection to hide and lazily load
internal submodules; and large projects such as NumPy and pandas maintain a
clear distinction between public APIs and internal cores. Although these designs
are typically described in ad hoc terms (e.g., ``wrapper modules'', ``lazy
imports'', ``internal cores'') they share a common structure: a stable
interface object that delegates to an opaque implementation. The Pythonic PIMPL
pattern gives this recurring structure a unified name and vocabulary.

\section{Discussion and Practical Considerations}
\label{sec:discussion}

Adapting the PIMPL idiom to Python raises several practical considerations. Although Python does not enforce strict separation between public and private components, the structural idea of keeping a small and stable interface while delegating to an internal implementation proves valuable in many contexts. A practitioner-oriented mapping from common design pressures to recommended Pythonic PIMPL strategies (and common pitfalls) is summarized in Table~\ref{tab:practitioner-cheatsheet}.

\begin{table*}[t]
\caption{Practitioner cheat sheet: goals, Pythonic PIMPL strategies, and pitfalls.}
\label{tab:practitioner-cheatsheet}
\small
\begin{tabular}{p{0.26\linewidth} p{0.38\linewidth} p{0.30\linewidth}}
\hline
\textbf{Goal/pressure} & \textbf{What to do} & \textbf{Pitfall and mitigation} \\
\hline
API stability under internal churn
& Keep a small stable interface; delegate behavior to \texttt{\_impl}
& Users may rely on \texttt{\_impl}; mitigate via docs, avoid reexporting internals, and tests that pin the public API \\

Dependency isolation/lazy imports
& Move heavy/optional dependencies into implementation/private modules; import lazily (method scope or module \texttt{\_\_getattr\_\_})
& Circular imports or surprising side effects; mitigate with deferred imports and clear import boundaries \\

Backend flexibility
& Select backend in constructor (i.e., \texttt{\_\_init\_\_})/factory; keep one public API
& Non-obvious behavior changes; mitigate with explicit configuration and documented selection rules \\

Reduced reachable internals
& Curate exports (\texttt{\_\_all\_\_}); keep implementations in private modules
& No true privacy; mitigate with conventions and tooling/linting \\

Testability
& Allow injecting/monkey patching/swapping implementations behind the interface; test the public contract
& Over-mocking internals; mitigate by focusing tests on public behavior \\

\hline
\end{tabular}
\end{table*}

One key consideration is establishing a clear boundary between the interface and the implementation classes. The interface should not contain substantial business logic, since mixing public and internal functionality reduces the benefits of the idiom. Instead, the interface should primarily forward method calls to the implementation, possibly performing light validation, argument normalization, or error handling. The implementation class, in turn, contains the actual operations, manages heavy dependencies, and encapsulates internal state.

Circular imports can arise when the interface and implementation classes are placed in separate modules. A not so common strategy is to import the implementation module inside the initializer of the interface-derived class or inside a factory function, instead of at the top of the file. This delays the import until it is needed and avoids module-level cycles. 

A PIMPL-style structure can significantly simplify internal evolution. For example, in one of our projects, an image processing component initially relied on pure Python loops. The implementation was later replaced with a NumPy-based vectorized version without altering the public interface. Because the public class delegated all work to an internal implementation stored in \texttt{self.\_impl}, the switch required no changes to user code. Later, additional implementations were introduced, including a GPU-accelerated backend selected dynamically according to input size or hardware availability.

Compared with other structural approaches, PIMPL-style composition offers advantages over inheritance. Using inheritance to separate interface from implementation often exposes internal methods in the public class and ties the two parts together more rigidly. Composition via an internal implementation object avoids these issues and keeps the public interface stable and uncluttered. Similarly, although Python permits monkey patching, using a clear interface--implementation separation leads to more maintainable and predictable codebases.

The idiom is not without limitations. Because Python lacks true private attributes, users can still access the internal implementation via attributes such as \texttt{obj.\_impl} if they choose to do so. The idiom therefore relies partly on convention and documentation. In addition, delegating calls through multiple layers introduces a small amount of overhead, although this is typically negligible compared with the cost of the computation being performed. For small or simple classes, introducing an implementation object may represent unnecessary complexity, and direct implementation within the public class may suffice.

Despite these limitations, many large Python systems stand to benefit from the PIMPL idiom. It provides a clear structure for managing internal change, reduces visible dependencies, and supports multiple implementations behind a consistent public API. These qualities make it a practical tool for improving the modularity and maintainability of evolving Python codebases. The key benefits and limitations of the idiom are summarized in Table~\ref{tab:pimpl-benefits-limitations}.

\begin{table*}[t]
\caption{Pythonic PIMPL: practitioner summary of benefits and limitations.}
\label{tab:pimpl-benefits-limitations}
\small
\setlength{\tabcolsep}{6pt}
\renewcommand{\arraystretch}{1.2}
\begin{tabular}{p{0.18\linewidth} p{0.76\linewidth}}
\hline
\textbf{Category} & \textbf{Item} \\
\hline

\multirow{5}{0.18\linewidth}{\textbf{Benefits}} &
\textbf{API stability under internal churn:} Refactor algorithms/data structures while preserving public entry points. \\
& \textbf{Dependency isolation and lazy imports:} Keep optional or heavy dependencies out of the import path until needed. \\
& \textbf{Backend flexibility:} Support multiple interchangeable implementations behind a single interface. \\
& \textbf{Reduced ``reachable internals'' risk:} Discourage users from depending on accidental internals by routing access through a curated interface. \\
& \textbf{Testability:} Swap implementations or inject mocks while testing the same public contract. \\
\hline

\multirow{3}{0.18\linewidth}{\textbf{Limitations}} &
\textbf{No true privacy:} \texttt{\_impl} remains accessible via introspection; the boundary is partly social or organizational. \\
& \textbf{Indirection overhead and debugging friction:} Stack traces and performance profiling may span delegation layers. \\
& \textbf{Over-engineering risk:} For small, internal-only components, the extra structure can reduce clarity rather than improve it. \\
\hline

\end{tabular}
\end{table*}

\section{Conclusion}
\label{sec:conclusion}

This paper has examined how the classical PIMPL idiom can be applied effectively in Python. This technique uses opaque delegation to implement a stable public interface backed by a changeable, hidden implementation. Although the idiom originated in C++ to address compilation dependencies and binary compatibility, its structural idea of separating a stable public interface from a changeable internal implementation maps naturally onto Python's object model. A Python class that stores an internal implementation object in an attribute such as \texttt{self.\_impl} can preserve a clear and consistent public API while allowing internal logic, dependencies, and performance strategies to evolve freely.

The use of PIMPL-style delegation in Python provides several practical advantages for large or long-lived libraries. Heavy or optional dependencies can be isolated inside the implementation object and imported only when needed. Multiple backend implementations can be selected at runtime without affecting the visible structure of the interface. Internal algorithms and data structures can be refactored or replaced without requiring changes to user code. These capabilities are increasingly important as Python becomes a foundation for scientific computing, data processing pipelines, and high-performance applications.

Although Python does not enforce strict separation between public and private components, adopting a PIMPL-inspired structure can help developers maintain cleaner boundaries, reduce clutter in public APIs, and limit unintended coupling between modules. By routing interactions through a small, stable interface and hiding volatile details in an internal implementation, the pattern encourages clients to depend on higher-level abstractions. In practice, this widens the dependency graph, localizes the impact of changes, and allows subsystems to evolve semi-independently behind a closed, implementation-specific namespace. The same separation also supports practical testing strategies in which implementations are swapped or mocked behind the interface.

Although the PIMPL idiom is already used informally in various
parts of the Python ecosystem, articulating how and why it applies
to Python offers value to practitioners seeking better modularity
and maintainability. The Pythonic PIMPL terminology provides a
useful conceptual framework for thinking about stable interfaces
and flexible internal structure in Python software. 

\subsection{Limitations and Future Work}
\label{sec:limitations}

This work is primarily conceptual and illustrative. We have not yet
conducted a systematic empirical study of how Python projects adopt
PIMPL-like structures over time, nor measured the impact of explicit
interface--implementation separation on maintenance costs, import-time
performance, or defect rates. Such studies would require sampling a
large corpus of real-world repositories, classifying their architectural
patterns, and tracking their evolution across versions.

Another limitation is that Python's lack of enforced privacy means that
the pattern relies on social conventions and documentation. Practitioners
may still reach into internal attributes such as \texttt{\_impl} when
debugging or experimenting, potentially undermining the intended
encapsulation. Tooling support (e.g., linters that flag access
to attributes documented as internal or that check imports against a declared
public API surface) could help reinforce the intended boundary.

Future work could proceed in several directions. One avenue is to carry out
empirical studies of PIMPL-like structures in major Python packages,
quantifying how often they arise and how they affect long-term evolution
in practice. A second direction is the design and evaluation of library
templates, refactoring assistants, or code generators that reduce the
boilerplate associated with interface/implementation delegation and make a
Pythonic PIMPL structure easier to adopt consistently. A third direction
involves integration with type checkers and IDEs, so that tools can
provide stronger support for distinguishing public and private APIs,
for navigating between interfaces and their implementations in large
codebases, and for warning when clients depend on undocumented internals.

\bibliographystyle{ACM-Reference-Format}
\bibliography{pimpl-python}

\end{document}